\documentclass[acmsmall,screen]{acmart}

\AtBeginDocument{%
  \providecommand\BibTeX{{%
    \normalfont B\kern-0.5em{\scshape i\kern-0.25em b}\kern-0.8em\TeX}}}

\setcopyright{acmcopyright}
\copyrightyear{2018}
\acmYear{2018}
\acmDOI{XXXXXXX.XXXXXXX}

\usepackage{multirow}
\usepackage{caption}
\usepackage{subcaption}
\usepackage{forest}
\usetikzlibrary{shadows}
\usepackage{soul}
\usepackage{enumitem}

\newcommand{\revise}[1]{\textcolor{black}{#1}}

\newcommand{\wps}[1]{whitepaper\xspace}

\renewcommand\footnotetextcopyrightpermission[1]{}

\renewcommand{\paragraph}[1]{\vskip 0.05in \noindent {\bf #1.}}

\def\tabref#1{Table~\ref{tab:#1}}
\def\figref#1{Figure~\ref{fig:#1}}

\begin{document}


\title{A Comprehensive Study of Governance Issues in Decentralized Finance Applications}

\author{Wei Ma}
\affiliation{%
  \institution{Nanyang Technology University}
  \country{Singapore}
}

\author{Chenguang Zhu}
\affiliation{%
  \institution{The University of Texas at Austin}
   \country{USA}
  }

\author{Ye Liu}
\affiliation{%
  \institution{Nanyang Technology University}
   \country{Singapore}
 }

\author{Xiaofei Xie}
\affiliation{%
  \institution{Singapore Management University}
   \country{Singapore}
}

\author{Yi Li}
\affiliation{%
 \institution{Nanyang Technology University}
  \country{Singapore}
}


\begin{abstract}
Decentralized Finance (DeFi) is a prominent application of smart contracts, representing a novel
financial paradigm in contrast to centralized finance. While DeFi applications are rapidly emerging
on mainstream blockchain platforms, their quality varies greatly, presenting numerous challenges,
particularly in terms of their governance mechanisms.
In this paper, we present a comprehensive study of governance issues in DeFi applications.
Drawing upon insights from industry reports and academic research articles, we develop a taxonomy
to categorize these governance issues.
We collect and build a dataset of 4,446 audit reports from 17
Web3 security companies, categorizing their governance issues according to our constructed
taxonomy. We conducted a thorough analysis of governance issues and identified vulnerabilities in
governance design and implementation, e.g., voting sybil attack and proposal front-running.
Our findings highlight a significant observation: the disparity between smart contract code and DeFi whitepapers plays a central role in these governance issues. 
As an initial step to address the challenges of code-whitepaper consistency checks for DeFi
applications, we built a machine-learning-based prototype, and validated its performance on eight
widely used DeFi projects, achieving a 56.14\% F1 score and a 80\% recall. The process of developing and evolving DeFi applications presents distinct traits that set it apart from conventional software systems. These include aspects like decentralization, transparency, and modification through governance mechanisms. Therefore, it's essential to explore whether adapting existing software development paradigms or creating a new methodology is more effective in constructing trustworthy DeFi applications.
Our study culminates in providing several
key practical implications for various DeFi stakeholders, including developers, users, researchers, and regulators, aiming to deepen the understanding of DeFi governance issues and contribute to the robust growth of DeFi systems. 
\end{abstract}


\maketitle

\section{Introduction}
Decentralized Finance~(DeFi)~\cite{zetzsche2020decentralized} has rapidly emerged as a transformative force in the financial world, challenging traditional parad-igms with its blockchain-based, intermediary-free model. The appeal of DeFi lies in its foundational principles: transparency, immutability, and openness, coupled with the anonymity it offers to users. At its core, DeFi contrasts starkly with centralized finance by empowering users with direct control over their transactions. This autonomy, facilitated by blockchain technology and smart contracts, marks a significant shift towards a more accessible and inclusive financial ecosystem. DeFi has also fueled a surge in innovation. This innovation is evident in the rapidly expanding DeFi landscape, encompassing varied financial services such as lending, trading, and asset management. A testament to its growing influence is the substantial capital influx, with billions of dollars currently locked in various DeFi protocols. A notable example is Uniswap, a decentralized cryptocurrency exchange boasting over 30 million active users and a total value locked (TVL) of approximately \$48.65 billion~\cite{statista}.

However, in addition to its immense opportunities, DeFi also faces a myriad of challenges~\cite{10.1145/3368089.3409740, KIRIMHAN2023113558, werner2022sok}.
One of the crucial aspects that require attention is governance~\cite{bhambhwani2022governing,
allen2020blockchain, beck2018governance, ekal2022defi, kiayias2022sok}.
Governance plays a central role in DeFi applications and serves as the foundation of
the DeFi ecosystem.
Effective governance in DeFi is vital for collective decision making, management of business models,
and the distribution of rewards within the ecosystem.
On the other hand, bad governance mechanisms may leave opportunities for unauthorized
manipulation of the DeFi protocol configurations and implementations, which may affect a large number of DeFi users.

In practice, DeFi governance faces many challenges.
First, some DeFi applications, such as Uniswap V1~\cite{uniswapv1}, operate without specific governance mechanisms, resulting in difficulties in maintaining and updating these applications.
This issue was rectified in its later versions, namely V2 and V3~\cite{uniswapv2}.
Second, vulnerable governance mechanisms expose an entire DeFi system to malicious attacks which
could result in massive financial losses.
Examples include the Beanstalk governance attack~\cite{daoattack} and the Build Finance DAO
incident~\cite{takeover}, which demonstrate vulnerabilities not just in code, but also in
governance designs.
Third, another serious issue that may harm DeFi users and investors is opaque or fraudulent
governance strategies.
It is a common practice for DeFi development teams to publish whitepapers about their projects in
advance, which documents the particular governance mechanisms to be adopted.
However, there are instances where the actual implementations deviate from these plans, sometimes
for the developers' own personal gains.
For example, CodeInc developers can issue tokens in excess of their declared amount and also have
the ability to destroy them to gain additional benefits, all without making any payment~\cite{coderinc}.
These discrepancies, while often subtle, can have a profound impact on the healthy growth of DeFi systems.
They raise questions regarding transparency and ethical practices in DeFi governance, potentially resulting
in diminished public trust.
A notable instance of such is the phenomenon of ``rug pulls''~\cite{10.1145/3623376}, where certain
DeFi teams (e.g., ARBIX FINANCE~\cite{cointelegraph}) exploit hidden governance loopholes to
rapidly deplete funds from the pool.

Robust and transparent governance mechanisms are crucial in the DeFi landscape.
They play a vital role in establishing trust among users and investors, which is key to encouraging long-term investment and
ensuring the sustainable development of the ecosystem.
However, there is a noticeable research gap in this area, particularly regarding the standardization of governance practices.
To address this, our paper embarks on a thorough examination of governance issues in DeFi applications, mainly by analyzing audit reports of influential DeFi applications.
Specifically, we delve into both academic research papers and high-quality industry reports to develop a comprehensive understanding of the DeFi governance taxonomy.
This taxonomy categorizes governance issues as well as assesses the issues based on their nature and severity, offering a detailed perspective on the present challenges in DeFi governance. We observe that 38\% of the high-severity issues identified pertain to governance, which is substantial.

Guided by the taxonomy we developed, an in-depth analysis of audit reports of various DeFi
applications revealed that the main governance issues center around the ownership structures and
incentive schemes.
For example, a recurring concern is the degree of centralization, especially 
in areas such as token distribution and protocol management, where a high-degree of centralization may endanger the application's credibility.
For example, critical privileges such as changing fee rates or transferring ownership are usually not expected to be centrally controlled.
In addition, our analysis also revealed that some serious governance issues in DeFi are rooted at the
mismatch between the proposed governance structures in whitepapers and their actual implementations. 
For instance, there are cases~\cite{hidden_mint} where DeFi owners reserve the right to mint any 
number of tokens by hiddening mint functions, which is never specified in the published governance 
structure.
These discrepancies indicate potential loopholes which may be exploited to bypass intended governance mechanisms.
As a first step towards automating the detection of these inconsistencies, we created a prototype tool based on large language model.
This tool is designed to identify these misalignments, aiming to improve the governance integrity in DeFi applications.

A crucial point to highlight from our study is that our findings offer several significant insights for multiple DeFi stakeholders. 
\revise{For researchers in software engineering, there is a need to study and develop governance frameworks and theories for DeFi, as well as methods for validating these governance systems. As an important application of blockchain, DeFi applications have a standardized governance system development process that is crucial. Unlike centralized software systems, the lifecycle management of DeFi diverges from traditional software approaches. In conventional software development, user needs and feedback drive continuous modifications. However, DeFi projects deviate from this model; their evolution is governed by their distinct governance systems. Stakeholders in DeFi should possess the authority to guide its evolution. Hence, it's imperative for software engineers to address challenges within DeFi governance and contribute towards establishing a comprehensive framework for its governance.}
For developers, it is essential to be cognizant of governance issues and to design transparent, ethical, and robust governance systems. For users and investors, they should investigate the governance structure of DeFi projects, such as the contract ownership, user privileges, and token power distribution. For regulators, it is crucial to both oversee governance structures and examine whitepapers of DeFi projects, as they are key to identify fradulent activities. 

These insights are instrumental in understanding and addressing the governance challenges within the DeFi sector. We aim to contribute to the development of robust and secure DeFi governance frameworks and foster a better understanding of governance issues, promote best practices, and facilitate sustainable growth of the DeFi ecosystem. In summary, we make the following contributions:

\begin{itemize}[itemsep=0.5em,topsep=0.5em,leftmargin=*]
    \item  \textit{A Taxonomy of DeFi Governance:} We created a detailed taxonomy for DeFi governance based on an extensive review of academic literature and industry reports. This taxonomy offers a structured approach to understanding and categorizing governance challenges in DeFi.
    \item \textit{A Comprehensive Analysis of DeFi Governance Issues:} Our paper provides a thorough examination of governance in DeFi applications. Through our analysis of audit reports by our governance taxonomy, we shed light on the present conditions, challenges, and unaddressed needs in DeFi governance. \revise{Our research highlights important issues that urgently need to be addressed in DeFi governance-related research.}
    \item \textit{An Inconsistency Checking Tool between DeFi Whitepapers and Implementations:} We have developed a new prototype tool employing large language model~(LLM). \revise{We use the text understanding ability and code ability of LLM to guess what possible variable names are possible to be used. Then, we use the symbolic expression matching the code and the expression from the whitepaper to verify if the financial model is correctly implemented. }
    This tool represents a pioneering effort to automatically identify and address 
    inconsistencies between the governance described in DeFi applications' whitepapers and their actual implementations, paving the way for more aligned and transparent governance practices.
\end{itemize}

 The paper is organized as follows.
Section~\ref{sec:method} introduces the three research questions under investigation, delineates our methodology, and details our data sources. Section~\ref{sec:findings} presents the results corresponding to each research question and discusses the implications of our study.
 Related studies to our work are reviewed in Section~\ref{sec:related_work}. Finally, Section~\ref{sec:threat} identifies potential threats and the corresponding mitigation strategies used. Our concluding remarks and a summary of the work are encapsulated in Section~\ref{sec:conclusion}.


\section{Methodology}
\label{sec:method}

\begin{figure}[]
  \centering
  \scalebox{0.6}{
  \includegraphics[width=\linewidth]{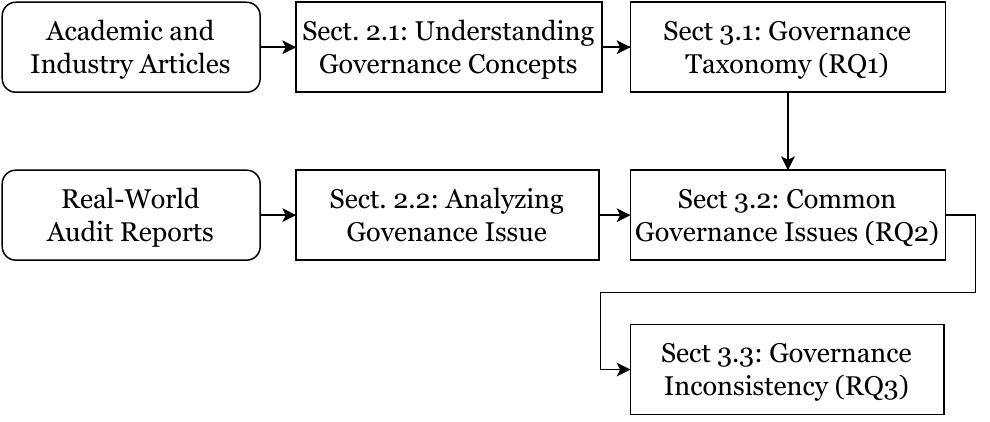}}
  \caption{Overview of our study methodology.}
  \label{fig:overview_method}
\end{figure}

\figref{overview_method} depicts the process that illustrates our analytical methodology and our three research questions in the different stages:
\begin{itemize}
    \item RQ1: What governance taxonomy can we use to analyze the governance issue?
    \item RQ2: What are the common governance issues in DeFi applications?
    \item RQ3: How closely do DeFi application developers follow the guidelines specified in whitepapers during application development? 
\end{itemize}

In our initial phase, we delve into the concept of DeFi governance as it is elucidated in academic literature and industry blogs. Our aim here is to develop a comprehensive governance taxonomy, a crucial framework that will underpin our entire analysis that guides us review the governance issues~(RQ1). Moving forward, we collect and scrutinize audit reports, employing our newly established governance taxonomy to dissect and understand governance issues in DeFi applications~(RQ2). A significant aspect of our study revolves around the role of DeFi development teams in designing and implementing governance mechanisms. Here, we examine the commitments outlined in whitepapers, recognizing these documents as not only informative tools but also as mediums through which trust is established with investors and users. 
Consequently, we also focus on assessing the extent to which development teams adhere to their designs and promises as articulated in their whitepapers~(RQ3).


\subsection{Understanding DeFi Governance Concepts} 
\label{sec:understanding_defi}




Governance, a pivotal concept in DeFi, lacks a universally accepted definition that encompasses its principles, scope, and role. This ambiguity poses a significant challenge in understanding and analyzing governance within DeFi. To address this crucial gap, we adopt the mapping study~\cite{PETERSEN20151}. \revise{We aimed at establishing a comprehensive taxonomy of DeFi governance.} This taxonomy is a classification system and a tool to dissect and categorize the multifaceted governance issues prevalent in DeFi applications. Through this endeavor, we aim to paint a clear picture of the current governance landscape in DeFi, highlighting key concerns and areas demanding further research attention.



\begin{figure}[]
	\centering
 \scalebox{0.7}{
	\includegraphics[width=0.8\textwidth]{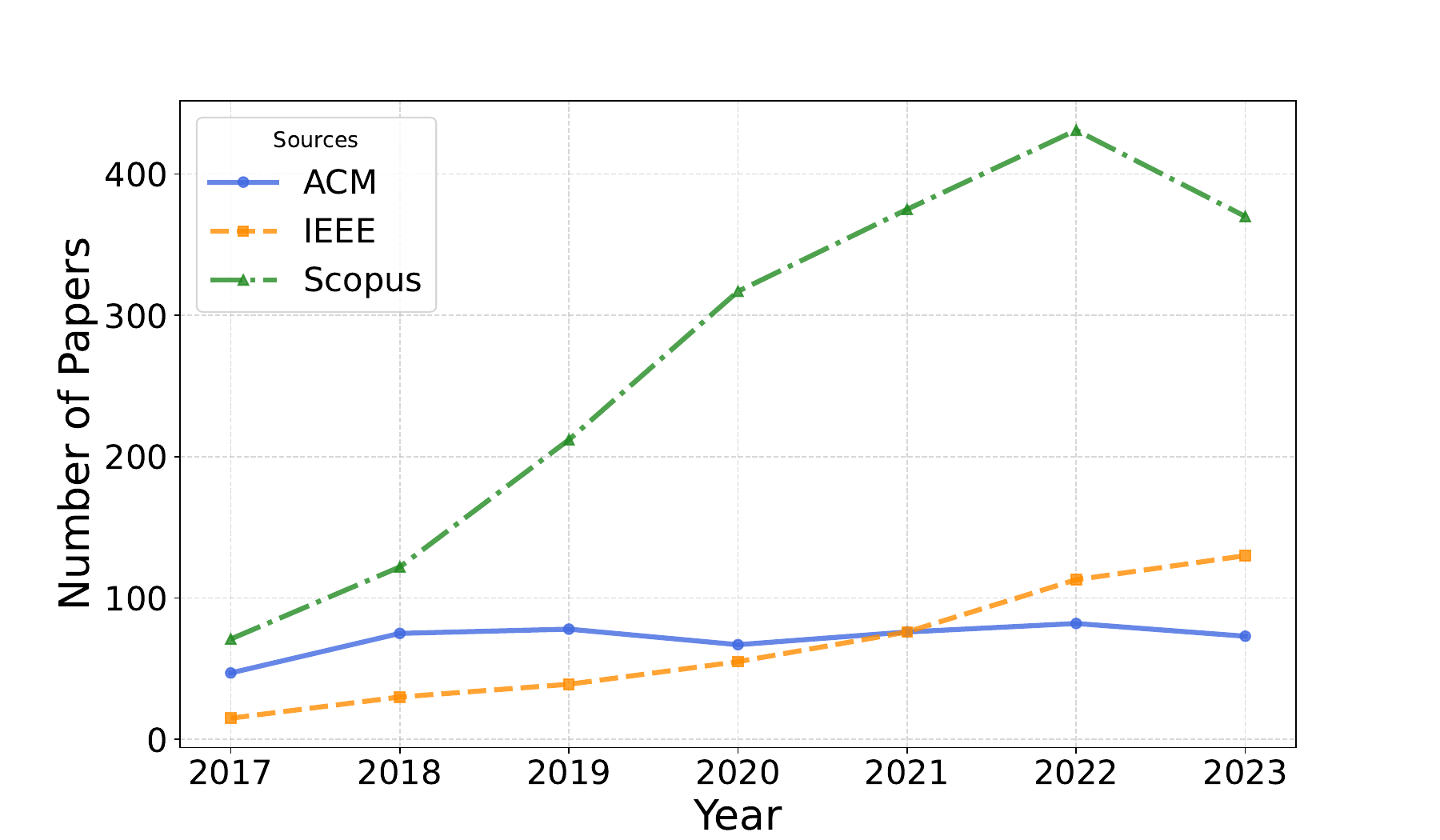}}
	\caption{Conference Paper Count by Year (2017-2023).}
	\label{fig:acm_ieee_scopus}
\end{figure}

We commenced our literature search with a systematic exploration of prominent databases, namely IEEE Xplore~\cite{ieee_xplore}, ACM Digital Library~\cite{acm_lib}, and Scopus~\cite{scopus}. We first defined the keywords that are directly related to the domain we were going to study, ``blockchain governance'', ``smart contract governance'', and ``DeFi governance''. These predefined keywords were used to search the related articles. 
This initial phase yielded a substantial corpus of articles: 458 from IEEE Xplore, 498 from ACM Digital Library, and 1898 from Scopus, all published between 2017 and 2023 as shown in \figref{acm_ieee_scopus}. \revise{It is evident that there is a growing emphasis on governance. It is important to acknowledge that the data for 2023 does not represent the entire year.} Recently, researchers have studied how GPT4 can be used to analyze documents~\cite{mellon2022does, shrestha2023we}. To distill this vast collection, we also employed GPT4 as a query assistant,
focusing specifically on the titles and abstracts to identify the 100 most pertinent articles from each database, with an emphasis on governance in decentralized finance applications. We upload the files that contain the title and abstracts to GPT4. Then, we instruct GPT4 to return the most related items with the prompt, "according to the 'title' and 'abstract', please use the topic model to extract the top 100 items from the file that related to the topic 'Decentralization Finance Application Governance' in a concise way."
Expanding our search horizon, we also utilized Google Scholar and Connected Papers~\cite{connectedpaper}, reviewing the first five pages of the search results to identify articles closely aligned with our research theme. We checked their titles and abstracts to see if they are related to decentralized finance and governance.
In the end, this comprehensive approach led to the selection of 34 academic articles. 

In our quest for a holistic understanding of DeFi governance, we extended our exploration beyond academic literature to include industry perspectives. Recognizing the critical insights that Web3 entities offer, we expanded our dataset to encompass 11 blog articles \revise{about DeFi governance} from leading firms~\cite{searchedlist} in this field, \revise{like OpenZeppelin}.
This inclusion of diverse, industry-specific perspectives ensures a more comprehensive understanding of DeFi governance. Notably, during searching articles, we discovered that several esteemed international organizations
have published reports on decentralization finance from EUROFI~\cite{eurofi}, WIFPR~\cite{wifpr},OECD~\cite{OECD},BIS~\cite{BIS} and Dutch Blockchain Coalition~\cite{smartgovernanceforsmartcontracts}.
We use Google engine and the aforementioned keywords to search for these reports. We consider only the internal organization or the national institute. This led us to incorporate five such reports into our dataset,
enriching our analysis with practical viewpoints. \revise{We used Google to collect industry blogs and organization reports, employing DeFi governance as a keyword search, and checked the first 10 pages.}
We have published the list of search results online~\cite{searchedlist}. 

In our quest to understand the specific aspects of governance emphasized in the literature, we recognize that a universally accepted definition of DeFi governance is still evolving~\cite{DURSUN2021102556,bhambhwani2023governing,doi:10.1080/10580530.2020.1720046,beck2018governance, bekemeier2021deceptive,Jensen2021AnIT,LIU2023111576,kiayias2022sok,wifpr}.
Key components identified in these studies include decision-making processes~\cite{DURSUN2021102556,beck2018governance,ekal2022defi,Sun2022DecentralizationII,LIU2023111576,schneider2020decentralized,BANINEMEH2023100127}, incentives for participation~\cite{beck2018governance,bhambhwani2022governing,Jensen2021AnIT,kiayias2022sok}, and issues related to ownership and decentralization~\cite{bhambhwani2022governing,Sun2022DecentralizationII,SANTANA2022121806}. A central aspect of governance in DeFi systems is the use of governance tokens, which often determine voting power in decision-making processes~\cite{beck2018governance,bhambhwani2022governing,Jensen2021AnIT,Sun2022DecentralizationII,kozhan_et_al:OASIcs.Tokenomics.2021.11}.
The distribution of these tokens and the mechanism by which they are used in governance, including both off-chain and on-chain methods~\cite{DURSUN2021102556,xu2023auto,ethereum, CertikGovernance, LIU2023111576}, are crucial aspects that shape the governance structure. 
Furthermore, participants in the governance process are often influenced by incentive models, which can include utility tokens, incentive models, and revenue models~\cite{pelt2021defining,bhambhwani2022governing,beck2018governance,bhambhwani2022governing,Jensen2021AnIT,kiayias2022sok,chaudhary2023interest} that are collectively called tokenomics.

An examination of governance from an industry perspective reveals notable congruences with academic viewpoints, especially in conceptualizing it as a rule-based framework for decision-making. While academic literature often delves into broad topics, such as the governance of foundational platforms like Ethereum, industry blogs~\cite{BinanceGovernance,BinanceTokenomics,BinanceRevenue,Certik,CertikGovernance,ethereum,messari,esatya} tend to diverge, placing a greater emphasis on the practical aspects of DeFi governance at the application layer. This includes a focus on the technical details, design strategies, and monitoring mechanisms of smart contract governance. These industry discussions typically revolve around several key themes, such as determining the scope of governance, identifying stakeholders, exploring different governance models, and conducting comparative analyses of their pros and cons. For instance, the intustries often discuss how governance mechanisms are integrated into the codebase, offering insights into real-world implementation challenges and successes. This practical orientation provides a complementary perspective to the theoretical frameworks discussed in academic circles. \revise{As a summary, the academy focuses on the nature of DeFi and its governace sysgtems with these aspects decision-making processes, incentive for participation, ownership and decentralization; while the industry emphasizes the mechanisms of implementation and the scenarios of real-world applications with the following views, design strategies, monitoring mechanisms and implementation choice.}


\subsection{Analyzing DeFi Governance in Audit Reports}

Audit reports, as products of expert scrutiny, are invaluable for studying the complex issues and challenges in DeFi governance. These reports, rich in high-quality data, provide detailed insights into smart contract issues that are critical to understanding DeFi governance. To harness this wealth of information, we gathered audit reports from a diverse range of reputable online sources, available in formats such as PDF and HTML. This varied collection ensured a comprehensive data pool.
Our analysis of these reports involved extracting and categorizing governance-related issues, thereby creating a detailed picture of the challenges and intricacies involved in DeFi governance.

\subsubsection{Data Resource.} 

In our pursuit of high-quality audit reports, we prioritized security companies known for their expertise and reliability, particularly focusing on Certik and OpenZeppelin, whose comprehensive reports are readily accessible on their websites. Recognizing that many security firms also publish their findings on GitHub, we employed a targeted search using the following keywords 
``audit'', ``audit report'', and ``smart contract audit'' to gather more reports. To ensure the credibility of these sources, we established a set of criteria: we examine an audit report only if the auditing company has over 1,000 Twitter followers—a threshold verified for authenticity using the tool FollowerAudit \cite{followeraudit}—or is recognized on the Etherscan directory of smart contract auditors~\cite{etherscanSecurity}. \revise{The website~\cite{etherscanSecurity} includes recognizable security companies, and audit reports from these companies have a higher chance of being good.}
\tabref{resource_audit_reports} lists 17 security companies, collectively contributing to a dataset of 4,446 audit reports. 



\begin{table*}[]
\centering
\caption{Commercial Security Company List as the Resource of Audit Reports.}
\scalebox{0.8}{
\begin{tabular}{|l|l|l|c|l|l|}
\hline
Company      & Official Website                               & \#Twitter Followers & Etherscan     & \#Reports & \#Issues       \\ \hline
Certik       & https://www.certik.com                         & 288,957           & \checkmark & 1,133 & 12,461\\ 
Openzeppelin & https://www.openzeppelin.com                   & 53,327            & \checkmark & 92 & 1,133\\ 
Immune Bytes & https://www.immunebytes.com                    & 738               & \checkmark & 83 & 673\\ 
Oak Security & https://www.oaksecurity.io                     & 1544              &            & 92 & 950\\ 
Cyberscope   & https://www.cyberscope.io                      & 12,703            &            & 336 & 2,364\\ 
Coinscope   & https://www.coinscope.co                      & 15,851           &            & 184 & 1,124\\ 
Solidified   & https://www.solidified.io                       & 2546              & \checkmark & 142 & 289 \\ 
HASHEX       & https://hashex.org                              & 10,472            & \checkmark & 30 & 280 \\ 
Zellic       & https://www.zellic.io                           & 6676              &            & 28  & 143\\ 
QuillAudits  & https://www.quillaudits.com                      & 12,563            & \checkmark & 85 & 512\\ 
CyStack      & https://cystack.net                             & 4643              & \checkmark & 11    & 48 \\ 
TechRate     & https://techrate.org                            & 12,466            & \checkmark & 1,745  & 3,049\\ 
Decentraland & https://decentraland.org                        & 631,806           &            & 1 & 3 \\ 
Chainsulting & https://chainsulting.de                        & 39,894            & \checkmark & 74 & 329\\ 
Somish       & https://www.somish.com                          & 349               & \checkmark & 9 & 113\\ 
PeckShield   & https://peckshield.com                          & 76,204            & \checkmark & 277   & 1,227\\ 
Quantstamp   & https://quantstamp.com                          & 79,148            & \checkmark & 124   & 1,339 \\  \hline
\multicolumn{1}{|l|}{Total}        &  \multicolumn{1}{c|}{-}                        &  -         & - & 4,446 & 26,037  \\ \hline
\end{tabular}}
\label{tab:resource_audit_reports}
\end{table*}

\subsubsection{Data Processing.} 
\begin{figure}[]
     \centering
     \begin{subfigure}[b]{0.48\textwidth}
         \centering
         \includegraphics[width=\textwidth]{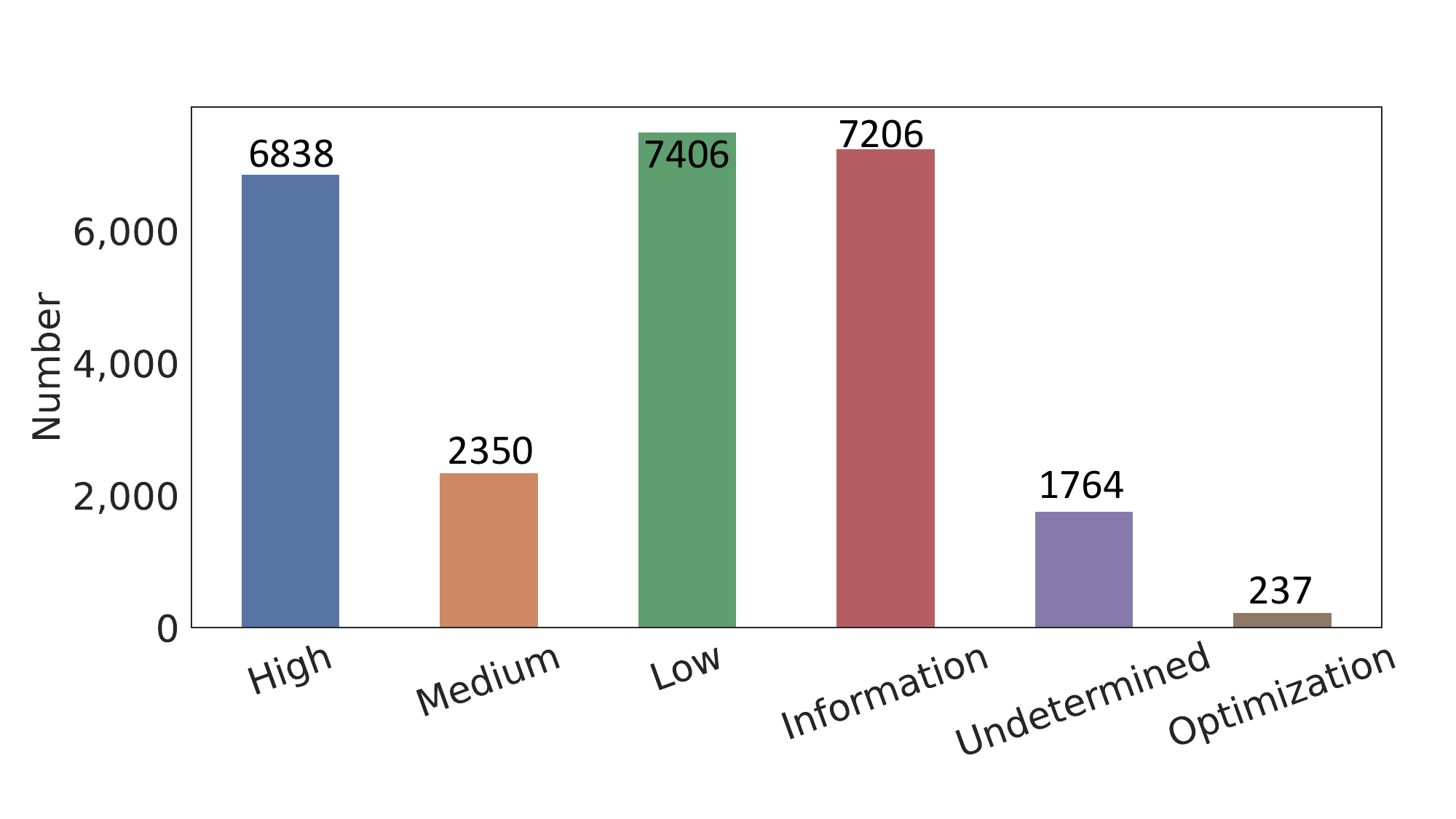}
          \vspace*{-2em}
         \caption{Distribution of Severity about all issues}
         \label{fig:severity_count}
     \end{subfigure} 
     \begin{subfigure}[b]{0.48\textwidth}
         \centering
         \includegraphics[width=\textwidth]{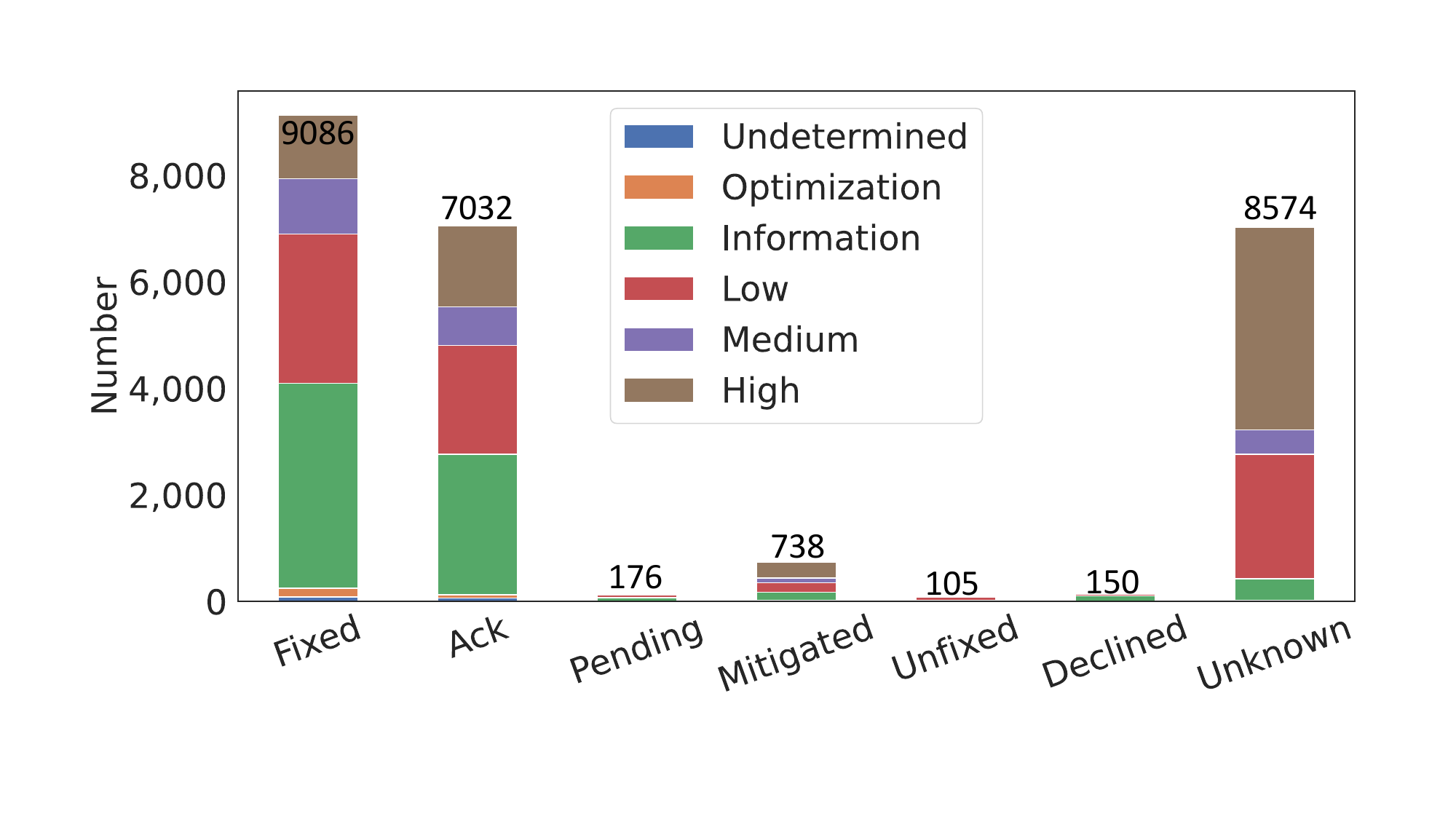}
          \vspace*{-2em}
         \caption{Distribution of Status about all issues.  }
         \label{fig:status_issues}
     \end{subfigure} 
     \vspace*{-1em}
     \caption{Severity and Status of the Issues in our dataset.}
             \vspace{-4mm}
\end{figure}

An audit report contains the issues found by experts in one project, and we need to extract each issue. We implemented a PDF parser and an HTML parser to convert raw audit reports into text format. Furthermore, \revise{we remove invalid text characters and then parse these texts into JSON format according to the different audit sections}, \textit{title}, \textit{severity}, \textit{recommendation}, \textit{status}, and \textit{description} of each issue.
The extraction leads to a total of 26,037 issues, as shown in the last column of \tabref{resource_audit_reports}. \figref{severity_count} illustrates the severity distribution of all issues. \revise{35\% of the issues are labeled  \textit{high} and \textit{medium} in terms of severity, which means the development team heavily depends on the audit to find the serious problems.}
\figref{status_issues} shows the resolution status distribution of all issues. 
We can conclude that most of the issues have been fixed~(34.89\%) or acknowledged~(27\%) by the developers. However, acknowledgment does not mean that the development team will fix these issues. \revise{If we compare the Fixed and Ack bars in \figref{status_issues}, many serious issues are acknowledged~(Ack) but not fixed. The reason behind it may be interesting but is beyond the scope of this work.}

We filter out non-governance issues and then group the remaining governance issues based on the built governance taxonomy.
To categorize the governance issues, we extracted keywords for each category in the governance taxonomy according to the collected articles, as shown in \tabref{keywords}. We also collect statistics on governance issues in terms of the status and severity of the issue. This can reflect what impact governance issues have and how developers handle them. Owing to the substantial amount of data acquired, manual analysis poses a significant workload. Consequently, we leverage Natural Language Processing (NLP) techniques for initial data processing. We employ BERTopic~\cite{grootendorst2022bertopic}
as an assistant to group the data and help us find common topics. We examine the clusters generated by BERTopic and determine the three most significant topics~(Top-3) by the frequency for each governance category.

\subsection{Investigating Consistency between DeFi Whitepapers and Implementations } 
\revise{During we analyze the governance issues, we find that some governance issues are related to the inconsistency between DeFi whitepapers and implementations. These inconsistencies were ignored in previous studies.} 
Inconsistencies between the whitepapers and the actual code can have a profound impact on investor trust and interest. Such discrepancies, even if they do not lead to vulnerabilities, can profoundly affect the credibility and success of a DeFi project. For users and investors alike, it is both interesting and vital to assess how faithfully developers adhere to their claims made in the whitepapers. This adherence is not just a matter of technical accuracy but also one of maintaining trust and transparency in the burgeoning world of decentralized finance.

To address the challenge of identifying governance issues related to whitepapers in DeFi projects, we employed an AI-based approach. Initially, we extracted relevant information by matching the two keywords 'whitepaper' and 'document' in audit reports. 
Recognizing the lack of existing tools for this specific purpose, we also developed a prototype tool that integrates the capabilities of ChatGPT~\cite{ma2023scope}. This tool is designed to facilitate semi-automated detection of discrepancies between whitepapers and source code. \revise{\figref{detection} illustrates our detector. We first ask ChatGPT to find the possible variable names that can appear in the code from the whitepaper. An expert will read the whitepaper and construct the financial expressions. Then, we build a tool to automatically extract the mathematical expression from the smart contract. Then, we use the symbolic expression to check if they are equal.} 

To validate our prototype tool, we collected the whitepapers and source codes from eight DeFi projects on the decentralization finance platform. \revise{The selected DeFi projects have to satisfy two criteria: 1). The whitepaper should contain the financial model and can be downloaded from the Internet; 2). The source code is public.} Our collection process revealed the often-overlooked challenge of accessing reliable sources; many projects, despite claims of openness, had invalid links due to various reasons, such as project discontinuation or poor maintenance.
We explore the ICO websites\footnote{https://icodrops.com/, https://icomarks.com/icos/defi} and the collected audit reports, and, in the end, we find eight projects whose source code and whitepaper are available.
A smart contract auditor \revise{that works on this field more than 1 years}
extracted the claims about tokenomics from these whitepapers, and then employed our tool to verify them in the code.




\section{Findings and Implications}
\label{sec:findings}


In this section, we delve into a detailed analysis of our three research questions based on our data and the analysis framework. We demonstrate our findings for each research question. In the end, we illustrate the implications. 

\subsection{Taxonomy of DeFi Governance~(RQ1)}

We identified their \textit{governance definitions} and \textit{what aspect they focus on DeFi governance}. After carefully reviewing these resources, we formulate a governance taxonomy specifically tailored to the domain of Decentralized Finance (DeFi). 
This taxonomy served as an analytical lens for our subsequent scrutiny of governance issues. 

\subsubsection{Developing a Taxonomy for DeFi Governance} 
\begin{figure}[]
	\centering
	\includegraphics[width=0.48\textwidth]{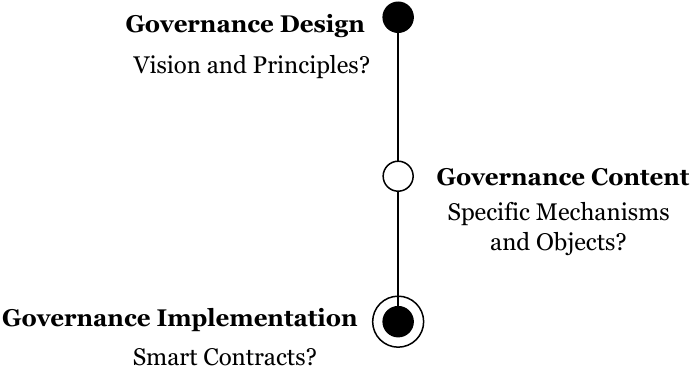}
	\caption{Three-Stage Development Process of DeFi Governance. }
	\label{fig:gov_hierarchy}
\end{figure}

The governance of one DeFi project should follow the typical software development pattern and is usually developed in a top-down manner, proceeding through three stages: \textit{governance design}, \textit{governance content} and \textit{governance implementation} as shown in \figref{gov_hierarchy}. 

\textbf{Governance Design} First, the DeFi team must establish a clear vision and principles guiding their governance approach to navigate the complexities of Decentralized Finance (DeFi) effectively. This fundamental step is crucial, as it shapes the subsequent decisions and strategies. 

\textbf{Governance Content} Next, the DeFi team should delve into the specifics of the governance structure. This involves selecting a suitable governance mechanism and justifying its appropriateness for their unique context. This step details the governance process's `what' and `why'
, ensuring all stakeholders understand the rationale behind the choices. 

\textbf{Governance Implementation} As the last step, the DeFi team implements the governance design. For a DeFi project to be deemed reliable, it should, at a minimum, detail its governance design in its whitepaper. 
An exemplary DeFi project not only makes claims in its whitepaper but also shows how these claims are realized in practical applications.

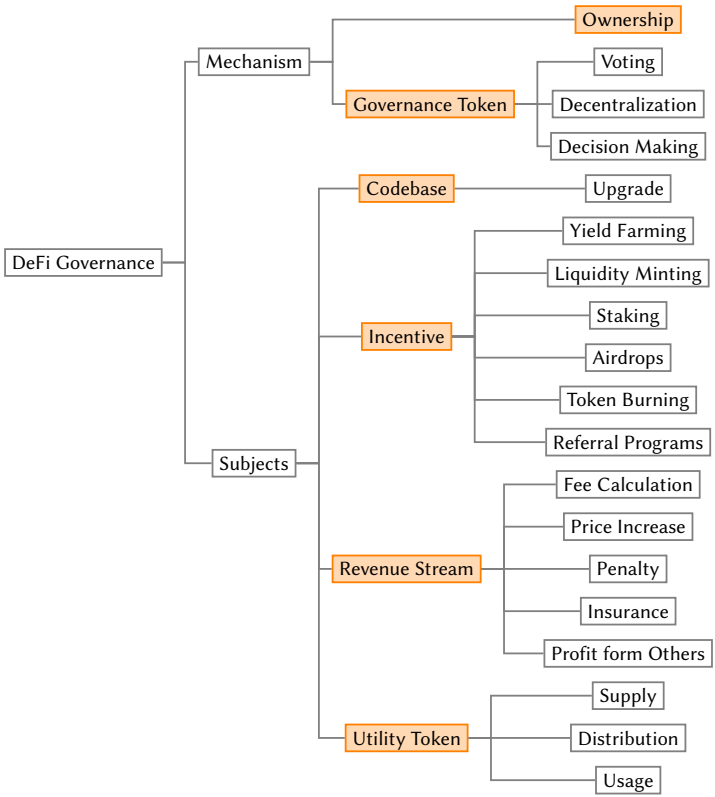
\begin{figure}[t]
\small
\scalebox{0.85}{
   \begin{forest}
        for tree={%
            my node,
            l sep+=5pt,
            grow'=east,
            edge={gray, thick},
            parent anchor=east,
            child anchor=west,
            if n children=0{tier=last}{},
            edge path={
                \noexpand\path [draw, \forestoption{edge}] (!u.parent anchor) -- +(10pt,0) |- (.child anchor)\forestoption{edge label};
            },
            if={isodd(n_children())}{
                for children={
                    if={equal(n,(n_children("!u")+1)/2)}{calign with current}{}
                }
            }{}
        }
        [DeFi Governance
        [Mechanism
        [Ownership, draw=orange, top color=orange!30, bottom color=orange!30,]
        [Governance Token, draw=orange, top color=orange!30, bottom color=orange!30,
        [Voting]
        [Decentralization]
        [Decision Making]
        ]
        ]
        [Subjects
        [Codebase, draw=orange, top color=orange!30, bottom color=orange!30,
        [Upgrade]]
        [Incentive, draw=orange, top color=orange!30, bottom color=orange!30,
        [Yield Farming]
        [Liquidity Minting]
        [Staking]
        [Airdrops]
        [Token Burning]
        [Referral Programs]]
        [Revenue Stream, draw=orange, top color=orange!30, bottom color=orange!30,
        [Fee Calculation]
        [Price Increase]
        [Penalty]
        [Insurance]
        [Profit form Others]
        ]
        [Utility Token, draw=orange, top color=orange!30, bottom color=orange!30,
        [Supply]
        [Distribution]
        [Usage]
        ]
        ]
        ]
\end{forest}}
\caption{The Overview of DeFi Governance Taxonomy.}
\label{fig:taxonomy}
\end{figure}

\figref{gov_hierarchy} illustrates the three steps to develop governance in practice. A good whitepaper should record the three steps and make sure that the governance is transparent to users and investors from the design to the end.
Based on our understanding of the extracted information of the collected articles as described in Section~\ref{sec:understanding_defi}
we developed our taxonomy to study governance issues. We approach DeFi governance as shown in \figref{taxonomy} from two perspectives: \textit{how governance} should be performed, and \textit{what should be governed}. According to the common content that we find from the collected articles, blogs, and reports, we used the six labels of governance issues with several related keywords that were highlighted in orange as shown in \figref{taxonomy}.
The leaf nodes list the common content of each category. \textit{How governance} is decided by the vision and objective of DeFi applications. 

Different types of DeFi applications have different usages and goals, so their governance design and the corresponding implementation are also different. For example, Uniswap 3 is a decentralized exchange platform and assigns voting power to users through liquidity mining\footnote{https://uniswap.org/governance}. Aave\footnote{https://aave.com/\#governance} is a lending platform and claims to be a decentralized non-custodial liquidity market protocol, so it use Decentralized Autonomous Organization~(DAO) to govern. Both are based on the on-chain governance but the later also need to communicate in the forum. \textit{What should be governed} describes the scope of DeFi governance. It consists of two parts, namely system mechanisms and underlying code. The DeFi project constructs an economic system that includes financial models, and all on-chain governance is conducted through code.

\textbf{How to govern} 
There are two types of governance mechanisms~\cite{ethereum, CertikGovernance, LIU2023111576}: off-chain and on-chain governance. Off-chain governance typically employs social approaches to reach a consensus for governance. On the other hand, on-chain governance utilizes coded mechanisms within the platform to achieve consensus. First, since our focus is on governance issues related to DeFi projects implemented in smart contracts, our taxonomy is based primarily on on-chain governance (specifically, governance tokens)~\cite{BinanceGovernance, messari, BARBEREAU2023102251, CertikGovernance, Certik}. The \textit{governance token} is used for decision making and is regarded as the power to propose and vote~\cite{DURSUN2021102556, bhambhwani2023governing,kiayias2022sok}. Usually, the governance token should be decentralized. Second, since the owner of a DeFi project often has certain privileges to govern smart contracts, \textit{ownership}~\cite{fusco2021decentralized,CertikGovernance} significantly influences the governance of DeFi projects and plays an important role in the governance mechanism. The right of belonging is a controversial topic and centralization does not comply with the Web3 manifesto, decentralization, but the actual situation may be more complicated. In the initial phase of the DeFi app, the team usually maintains ownership for convenience of updating. Although excessive rights are considered risky, when analyzing data, we have observed that the presence of an ower role is needed in emergency situations, e.g., stopping an attacking transaction. The developer team should justify why they keep ownership, what power the owner role has, and how they manage the owner role.

\textbf{What should be governed} We identified two key areas: tokenomics~\cite{pelt2021defining,bhambhwani2022governing,beck2018governance,bhambhwani2022governing,Jensen2021AnIT,kiayias2022sok,chaudhary2023interest,BinanceTokenomics,BinanceRevenue,fusco2021decentralized,werner2020towards} and codebase~\cite{werner2020towards}. \textit{Tokenomics} refers to the ecosystem defined by DeFi projects and comprises three essential components: 1) Utility tokens~\cite{BinanceTokenomics}, which serve as proof of access to DeFi services such as payments, staking, and lending. The supply of Utilit token usually has the maximum limitation. The initial distribution of tokens greatly affects the interest allocation, the security and reliability of the project.  2) Revenue streams~\cite{BinanceRevenue,fusco2021decentralized}, which outlines how DeFi projects generate profits. The revenue mechanism directly affects the survival time of the project and is a key issue to focus on in the DeFi governance. Revenue streams involve charging fees from users, increasing token prices, insurance, and the profits from other projects. 3) Incentive mechanisms~\cite{BinanceTokenomics, fusco2021decentralized}, which detail how participants are incentivized to ensure long-term sustainability of the DeFi project. Depending on the target of DeFi applications, there are various incentive mechanisms that reward participants. Yield farming is to earn the rewards by locking the token. Liquidity mining rewards the liquidity providers. Other incentive mechanisms include staking, airdrops, token burning, and referral programs.

\textit{Codebase} pertains to the implementation of DeFi applications. Governance of the codebase involves determining how to update the implementation and rectify code-related issues, e.g., fixing vulnerabilites or code optimization. These modifications directly influence the DeFi project's functionality, impacting every user. Hence, regulating changes in the code is a crucial aspect of contract governance. In this study, we focus on the issues of code updating, that is, how DeFi applications maintain their code.

\tikzset{
        my node/.style={
            draw=gray,
            inner color=white,
            outer color=white,
            thick,
            minimum width=1cm,
            text height=1.5ex,
            text depth=0ex,
            font=\sffamily,
        }
    }



\subsection{Common Governance Issues~(RQ2)}
\subsubsection{Distribution of Governance Issues} 

\begin{figure}[]
	\centering
 \scalebox{0.65}{
    \includegraphics[width=\linewidth]{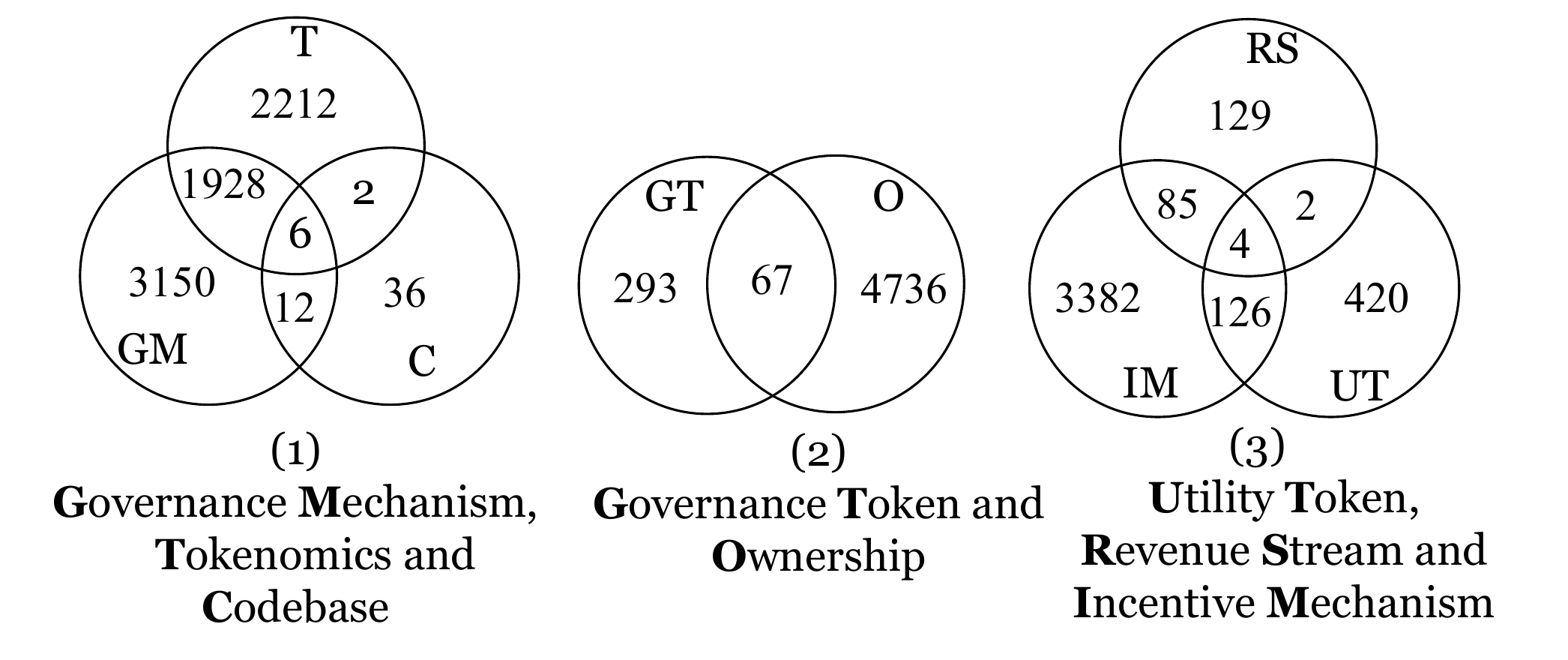}}
	\caption{ Overlapping of Different Categories about Governance Issues.}
	\label{fig:overlapping_gov}
\end{figure}

Initially, we must filter out non-government issues. \tabref{keywords} presents the raw keywords we originally used for issue classification, aimed at extracting governance issues based on our established governance taxonomy. Our taxonomy contains six class labels, as depicted in the subcategory column. These labels are grouped into three categories: governance mechanism, tokenomics, and codebase. These keywords were derived from the papers and blogs we collected. We eliminated the issues that did not include these keywords, yielding a total of 7,346 governance issues. The status and severity of the governance issues for each category are demonstrated, respectively, in \tabref{status_gov_issues} and \tabref{severity_gov_issues}. The final columns in \tabref{status_gov_issues} and \tabref{severity_gov_issues} reveal that most of the governance issues relate to ownership and incentive mechanisms. Discounting the status-unknown issues, it is evident that almost all status-known governance issues are addressed by the development team, as seen in \tabref{status_gov_issues}. \tabref{severity_gov_issues} indicates that most governance issues are labeled with high or medium severity. Compared to \figref{severity_count}, it is clear that a significant number of high-severity issues are governance related~(about 38\%). \figref{overlapping_gov} illustrates the extent of overlap among different categories of governance issues. It is evident from \figref{overlapping_gov} that the majority of these issues are unique to their respective categories and do not overlap.

\begin{table*}[]
\centering
\caption{Raw Keywords for Issue Classification.}
\vspace*{-4mm}
\large
\resizebox{0.9\textwidth}{!}{
\begin{tabular}{|p{3cm}|l|p{6cm}|c|}
\hline
Category                              & Subcategory        & Keywords                                                                 & Citations                               \\ \hline
\multirow{2}{*}{ \begin{tabular}[c]{@{}l@{}}\\Governance\\ Mechanism\end{tabular} } & \multirow{2}{*}{ \begin{tabular}[c]{@{}l@{}} \\ Governance Token   \end{tabular} }  & governance token, vote,  proposal, deceision-making, tally, abstation, quorum, veto      &       \multirow{2}{*}{ \begin{tabular}[c]{@{}l@{}}\\ \cite{DURSUN2021102556,beck2018governance,ekal2022defi,Sun2022DecentralizationII,LIU2023111576,schneider2020decentralized}\\ \cite{BANINEMEH2023100127, bhambhwani2022governing,Sun2022DecentralizationII,SANTANA2022121806,jensen2021decentralized} \end{tabular} }      \\ \cline{2-3} 
                                      & Ownership           & owner, ownership, privilege                       &                                                      \\ \hline
\multirow{5}{*}{\begin{tabular}[c]{@{}l@{}}\\ \\ \\ \\Tokenomics\end{tabular}}           & \multirow{2}{*}{ \begin{tabular}[c]{@{}l@{}} Utility Token   \end{tabular} }      & supply, token distribution, token name, token usage, asset token, token utility          &               \multirow{5}{*}{\begin{tabular}[c]{@{}l@{}}\\ \\ \\ \\ \cite{pelt2021defining,bhambhwani2022governing,beck2018governance,chaudhary2023interest,Jensen2021AnIT,kiayias2022sok}\\ \cite{balcerzak2022blockchain, kaal2020blockchain, hunhevicz2022applications, fiorentino2021blockchain, beck2018governance}\end{tabular}}                          \\ \cline{2-3} 
                                      & \multirow{2}{*}{ \begin{tabular}[c]{@{}l@{}}\\ \\ Revenue Stream \end{tabular} }  & transaction fee, trading fee, marketplace fee, borrow rate, protocol fee, premium fee, performance fee, token issuance, generic fee, interest rate, charge a fee & \\ 
                                      \cline{2-3} 
                                      &  \multirow{2}{*}{ \begin{tabular}[c]{@{}l@{}}\\ Incentive Mechanism\end{tabular} }  &  lock up, total value locked, yield, borrow, airdrop, burn, stake, liquidity, lend, loan, referral, mint, incentive & \\ 
                                      \hline
Codebase                              & Code                & update contract, upgradable                                                                   &      \cite{smartgovernanceforsmartcontracts,openzeppelin,ethereum1,oceanprotocol,zeppelinos,polygon}    \\ \hline
\end{tabular}
}
\label{tab:keywords}
\end{table*}



We conclude \textit{the following points}: 1) ownership and incentive mechanisms are two of the most common governance issues; 2) apart from governance issues with unknown status, most governance problems have been resolved or acknowledged; 3) the number of high-severity issues is the highest among all severity levels of governance problems; 4) a significant number of high-severity issues are governance related~(about 38\%); 5) there is a small overlap among these governance categories.
\begin{table}[]
\centering
\caption{Status of Governance Issues. The number in the last row are counted by excluding the overlapping.}
\label{tab:status_gov_issues}
\scalebox{0.8}{
\begin{tabular}{|l|c|c|c|c|c|c|c|c|c|c|}
\hline
Category          & Fixed & Ack  & Mitigated & Pending & Unfixed & Declined & Unknown & All \\ \hline
Gov Token     & \textbf{175}   & \textbf{118}  & 19        & 1       & 3       & 1        & 43      &   360    \\ 
Ownership     & \textbf{688}   & \textbf{1372} & 214       & 23      & 9       & 8        & 2,489    &   4803 \\ \hline
Utility Token & \textbf{115}   & \textbf{273}  & 40        & 6       & 3       & 2        & 113     &   552 \\ 
Revenue       & \textbf{88}    & \textbf{80}   & 8         & 1       & 2       & 3        & 38      &   220 \\ 
Incentive     & \textbf{879}   & \textbf{986}  & 130       & 16      & 12      & 9        & 1,565    &   3,597 \\ \hline
Codebase      & \textbf{20}    & \textbf{14}  & 7         & 0       & 1       & 0        & 14      &   56  \\ \hline
Total     & \textbf{1,689}  & \textbf{2,303}  & 325   &  41   &  26    &   17   &  2,945   &   \textit{7,346}    \\ \hline
\end{tabular}}
\end{table}

\begin{table}[]
\centering
\caption{Severity of Governance Issues. The number in the last row are counted by excluding the overlapping.}
\label{tab:severity_gov_issues}
\scalebox{0.8}{
\begin{tabular}{|l|c|c|c|c|c|c|c|}
\hline
Category          & High & Medium & Low & Information & Optim & Undet & All \\ \hline
Gov Token     & 101  & 77     & \textbf{130} & 44          & 0            & 8    & 360 \\ 
Ownership     & \textbf{1,814} & 447    & 637 & 327         & 3            & 1575    & 4,803 \\ \hline
Utility Token & \textbf{279}  & 98     & 118 & 48          & 0            & 9     & 552 \\ 
Revenue       & 46   & 43     & \textbf{82}  & 41          & 0            & 8     &  220 \\ 
Incentive     & \textbf{972}  & 421    & 832 & 356         & 3            & 1013    & 3,597\\ \hline
Codebase      & \textbf{21}   & 14     & 14  & 5           & 1            & 1    & 56 \\ \hline
Total      &  \textbf{2606}   & 866   & 1,554 &  713      &   7    &   1600   & \textit{7,346} \\ \hline
\end{tabular}}
\end{table}


\subsubsection{Hierarchy of Governance Vulnerability} 
\begin{figure}[]
	\centering
	\includegraphics[width=0.6\textwidth]{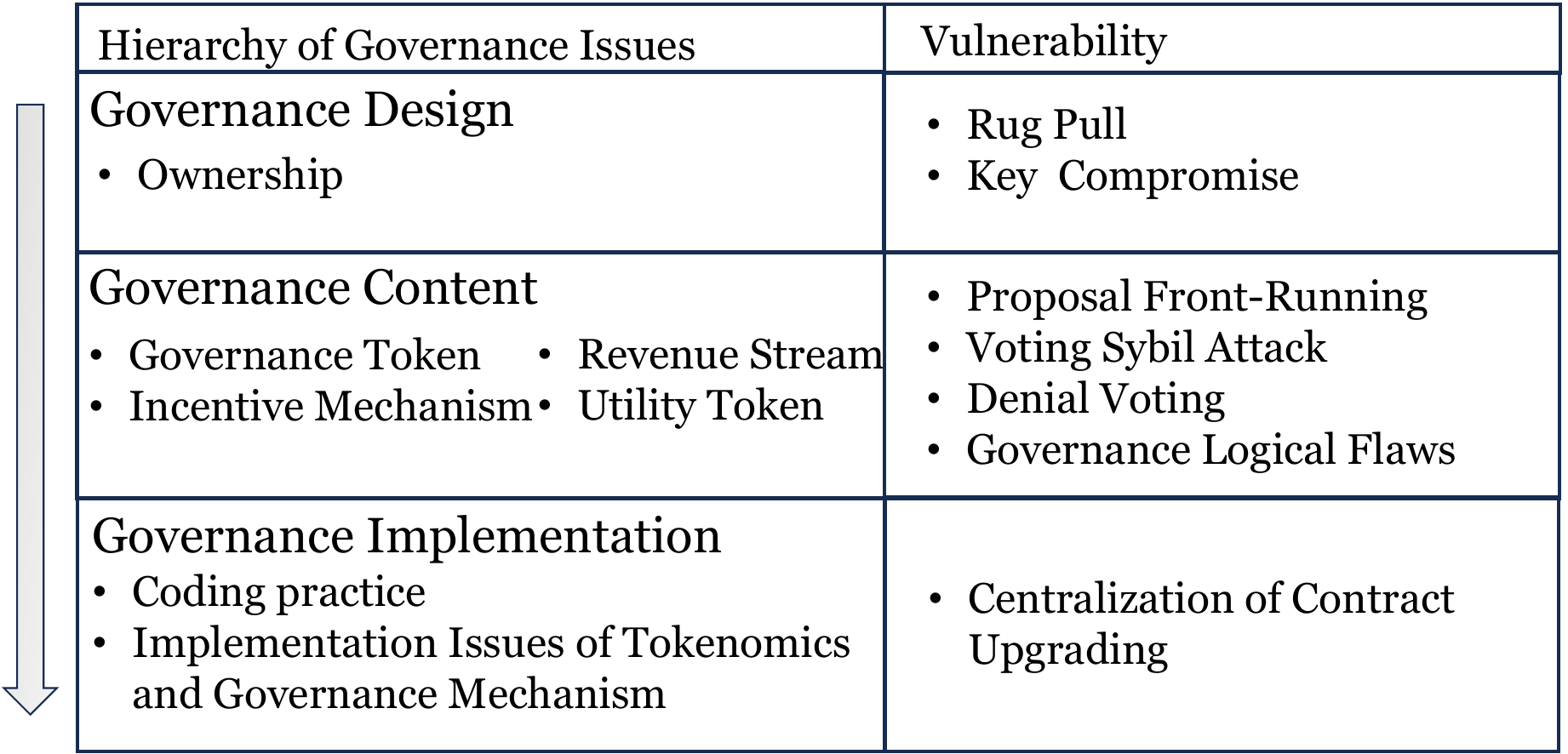}
	\caption{Hierarchy of Governance Issue.}
	\label{fig:hierarchy_issues}
\end{figure}

\revise{To delve deeper into governance issues and their related vulnerability, we have categorized these issues based on the design and implementation process of governance, as depicted in \figref{hierarchy_issues}. The first column in the figure outlines the governance issues at each developmental stage, while the second column identifies the corresponding vulnerabilities. For each stage, we summarized vulnerabilities by identifying key security-related terms, such as "attack", and then conducted a manual review of the filtered issues. Initially, in the governance design stage, ownership is determined based on the vision and principles of the DeFi project. Ownership is the core of DeFi governance. It decides which governance mechanism will be applied. For example, if ownership is decided to belong to all participants, it should use the DAO~(decentralized autonomous organization) governance model. During this stage, the owner rights should be carefully designed. Flaws in ownership design can lead to risks like 'rug-pulling'. Owners, typically vested with extensive rights to manipulate contracts and maintain privileged functions, can easily misappropriate funds from users. The special rights associated with governance roles can, if compromised, pose a threat to the entire DeFi application. }

\revise{
In the subsequent stage, issues concerning governance tokens, revenue streams, incentive mechanisms, and utility tokens emerge, all of which contribute to the formation of governance content. This stage often brings to light governance functionality issues, such as proposal front-running, Sybil attacks in voting, denial voting, and inconsistencies in the governance process relative to its design. For instance, in proposal front-running, attackers can prematurely reach a consensus on a malicious proposal~\cite{OneSwap}. In voting Sybil attacks~\cite{LuckyChip}, an attacker might rapidly acquire substantial voting weight through flash loans. Denial voting often results from the depletion of gas fees~\cite{DerivaDEX}. Governance process flaws can significantly increase the susceptibility to attacks. For example, DerivaDEX's governance structure and its use of the Diamond proxy pattern are vulnerable to exploitation by malicious actors~\cite{DerivaDEX}. The implementation of governance may be different from the designed documents, introducing some risks such as phishing attacks~\cite{openzeppelin_audit} or causing the fairness problem~\cite{Notable_audit} to the DeFi projects.}

\revise{
Lastly, the governance implementation stage involves addressing issues related to coding practices and the implementation of tokenomics and governance mechanisms. The way governance roles manage and update the DeFi system is crucial. Incorrect initialization~\cite{CrowdSwap} or inappropriate upgrades can precipitate the failure of a DeFi project~\cite{Tokensfarm}.}

\subsubsection{Common Topics on Governance Issues} We use BERTopic~\cite{grootendorst2022bertopic} to cluster governance issues for each category. \tabref{top_3_topics_all} demonstrates the top-3 frequent topic words, respectively. Based on the clusters in \tabref{top_3_topics_all}, we make the following summary:

\paragraph{Governance Token} The first is related to the management of the proposal. It includes the expiration of the proposal, unexpected cancellations, and non-unique keys to identify voting topics. The second one is about the voting process, especially the management of voting power, i.e., how it is transferred, burned, or minted. The third is about the decision-making processes and the governance vulnerabilities, e.g., the delay of executing decisions and malicious proposals.  

\paragraph{Ownership} The first is about the centralization risk that the owners are authorized too much or less, e.g., stopping any transaction or lacking the handling-emergency right. The second is about the administration key management; the admin roles are assigned with some privileged functions and the leak of admin keys must result in huge loss. The third is about the management of the blacklist.

\paragraph{Utility Token} The first concerns the initial token distribution. For example, deployers distribute tokens without obtaining consensus from the community, and few whale accounts can hold almost all tokens. The second is about the token total supply. For example, the total token supply is not consistent with the whitepaper, and the total supply can be increased or decreased without any restriction. The third is about the usage problems of the utility token, e.g., price settings, liquidity, and reward calculation.

\paragraph{Incentive Mechanism} The first is the centralization risks and the privilege function, where owners have the authority to manipulate functions related to the incentive mechanism, such as modifying the fee rate. The second is about the minting function, including the minting restriction and limitation, minting authority, and access control. The third is about the administration key management; the improper key management can resulting in the incentive mechanism can be manipulated by the third party.

\paragraph{Revenue Stream} The first concerns fee configuration issues, e.g., fee calculation, transaction fee manipulation, and inconsistent fee setting. The second is about borrowing and loaning issues, e.g., the design of borrowing and loaning processes or the interest rate configurations. The third is about the incompatibility between the non-deflationary token and the deflationary tokens. 

\paragraph{Codebase} The first major concern is the technical aspect, such as improper initialization or incorrect implementation during upgrades. These issues can lead to catastrophic update failures or expose vulnerabilities in the contract. The second concern is the risk of centralization in contract upgradeability. If an attacker gains control of the 'owner' role, they could update the contract with malicious intent, potentially leading to substantial losses. 

\begin{table*}[]
\centering
\caption{Top-3 Topic Words in all governance issues.}
\label{tab:top_3_topics_all}
\vspace{-1em}
\resizebox{0.9\textwidth}{!}{
\begin{tabular}{|l|p{4.5cm}|p{4.5cm}|p{4.5cm}|}
\hline
\textbf{Category}      & \textbf{Top 1}                          & \textbf{Top 2}                                          & \textbf{Top 3}                                       \\ \hline
Gov Token     & proposal, contract, token      & voting power, power moved                      & governance, macilious, lack event           \\ \hline
Ownership     & centralization risk            & trust issue admin, keys                        & blacklisted contract                        \\ \hline
Utility Token & token distribution initial     & total supply, wrong total, supply restriction  & rewards, price, users, duplicate, liquidity \\ \hline
Incentive     & owner privileges, centralization risk         &    mint token          &    trust issue admin, keys              \\ \hline
Revenue       & protocol fee, transaction      & interest rate, borrow, manipulate              & incompatibility deationary tokens           \\ \hline
Codebase      & upgradeable contracts, storage & centralized control contract, contract upgrade & -                                           \\ \hline
\end{tabular}}
\end{table*}

\subsection{DeFi Whitepaper-Implementation Inconsistency~(RQ3)}

 \subsubsection{Governance Inconsistency Issue} 
 We used two keywords, ``whitepaper'' and ``document'' to filter the governance issues and find \textit{136} issues related to the governance design. We use the BERTopic~\cite{grootendorst2022bertopic} model to analyze the main topics in these \textbf{136} issues and find that the inconsistency between the document and implementation is the main concern~(top 1). These inconsistencies usually have three cases; a) the code does not implement the features in the whitepaper; b) the code implements features that are not described in the whitepaper; c) the code implementation mismatches the design in the whitepaper, e.g., the tokenomics configurations like fee and interest rate.

 \subsubsection{Detection of Inconsistency}
\begin{figure}[]
	\centering
 \scalebox{0.8}{
	\includegraphics[width=\textwidth]{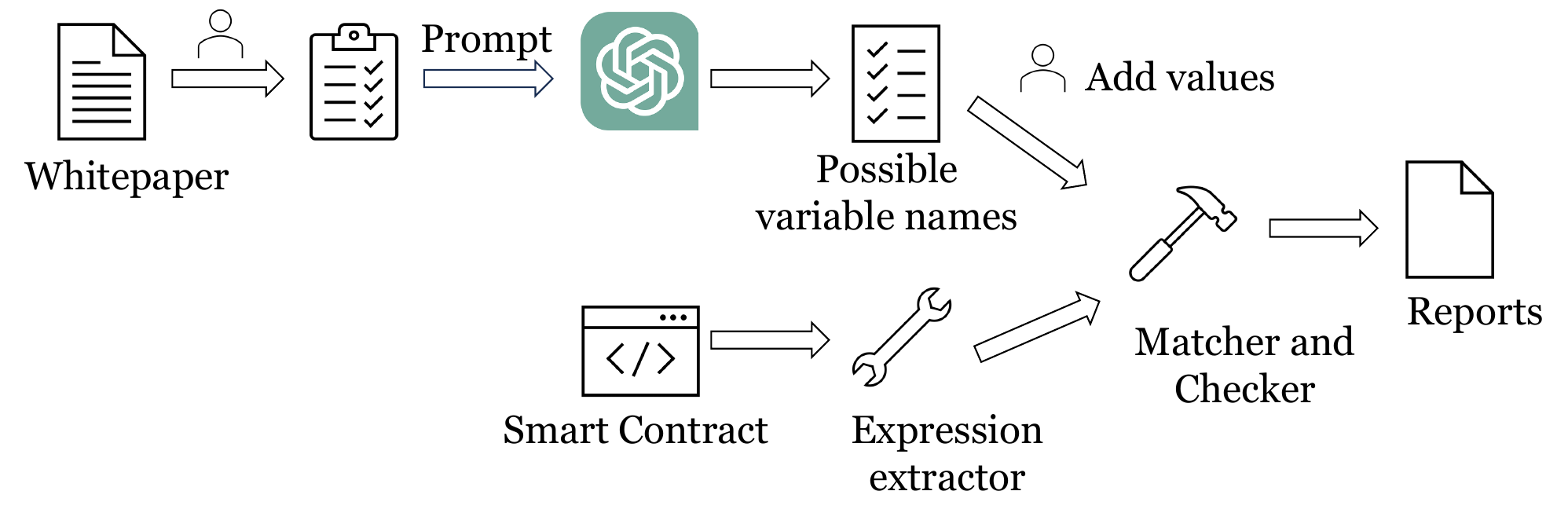}}
	\caption{Inconsistency Detection of Parameters about Tokenomics.}
	\label{fig:detection}
\end{figure}

 There are many difficulties in detecting the inconsistency between the whitepaper and the implementation. Whitepapers from different developers have different formats and contain a variety of information. The source codes from the different developers also have different styles. These factors make matching between the whitepaper and the code difficult. To study whether it is possible to detect these inconsistencies automatically or semi-automatically, we make a prototype tool as the first attempt. This tool checks if the fee setting is consistent between the whiepaper and the implementation. \figref{detection} demonstrates the overview of our prototype tool. We use the programming ability~\cite{biswas2023role} of ChatGPT to obtain possible variable names of the DeFi configurations in the code. We extract the tokenomics configuration settings from the whitepaper and then check if they match the code. \tabref{inconsistency_check} shows the results with the F1 score of 56.14\% and recall of 80\% on our collected DeFi projects. The second column shows the number of tokenomics configuration settings in the whitepaper. The predicted P/N in the third column means the predicted positive/negative by our tool, where we label the inconsistency as 1 and the consistency as 0.  FP and FN mean false positive and false negative. Our tool is available online.\footnote{\url{https://anonymous.4open.science/r/consistency\_checker-CD88}}
 
\begin{table}[]
\centering
\caption{DeFi Tokenomics Configuration Inconsistency Checking, and F1 = 56.14\%.}
\label{tab:inconsistency_check}
\scalebox{0.8}{
\begin{tabular}{|l|c|c|c|c|}
\hline
             & \multicolumn{1}{l|}{No. of Params} & Predicted P/N & \multicolumn{1}{l|}{FP} & \multicolumn{1}{l|}{FN} \\ \hline
MoonGame~\cite{MoonGame}    & 5                                     & 0/5     & 0                       & 0                       \\ 
Panther-Farm~\cite{PantherSwap} & 6                                     & 4/2     & 2                       & 1                       \\ 
HFTToken~\cite{hashflow}     & 6                                     & 5/1     & 3                       & 0                       \\ 
ShivaToken~\cite{SHIVA}   & 8                                     & 3/5     & 3                       & 2                       \\ 
ASENIX~\cite{ASENIX}       & 6                                     & 4/2     & 3                       & 0                       \\ 
Biokript~\cite{biokript}     & 7                                     & 7/0     & 1                       & 0                       \\ 
Bitchmdefi~\cite{Bitchemicaldefi}   & 7                                     & 6/1     & 1                       & 1                       \\ 
Eteru~\cite{eternumland}        & 9                                     & 8/1     & 8                       & 0                       \\ \hline
Total    & 54                 & 37/17    & 21    & 4   \\ \hline
\end{tabular}}
\end{table}


\subsection{Implications}



Our findings from this study derive multiple implications for various stakeholders in the DeFi ecosystem, including researchers, DeFi developers, investors, users, policymakers, and regulatory bodies. Specifically, we provide insights on developing and improving governance frameworks in DeFi projects, highlight the challenges and vulnerabilities in DeFi governance, and facilitate the formulation of guidelines and best practices in this domain.

\paragraph{For Researchers} 
\textit{First}, in the realm of software engineering (SE) research, there is a pressing need to delve into DeFi governance frameworks. Governance challenges form a substantial part of the hurdles DeFi applications face, necessitating thorough research and solutions. Our taxonomy serves as a foundation for this exploration. Yet, numerous challenges remain unaddressed. For instance, the absence of a governance development model akin to software development frameworks hampers the design and implementation of effective governance strategies. Furthermore, the fairness in tokenomics and the balancing act in centralization demand in-depth study. For example, determining the extent of control for 'owners' and establishing ownership criteria are critical. In certain scenarios, centralization can safeguard a DeFi project, such as halting a malicious, unvalidated transaction. A deeper understanding and resolution of these issues could significantly enhance investor confidence and the success of DeFi applications.
\textit{Second}, a robust verification methodology is vital for scrutinizing the governance system design prior to implementation. Given that smart contracts are immutable post-deployment, rectifying defects is not as straightforward as in traditional software systems. From proposal initiation to final execution, each state transition within the governance process warrants careful verification. Overlooking flaws in governance design can have dire consequences, such as losing control to hackers. Addressing potential flaws proactively is crucial, as fixing high-level issues post-deployment in DeFi systems can be exceedingly costly.
\textit{Third}, ensuring consistency between DeFi whitepapers and their implementation is paramount. This alignment is crucial for transparent and accurate communication with investors and users.  Developing semi- or fully automatic approaches to verify this consistency is an essential step forward. The prototype tool developed in our research could pave the way for more advanced systems capable of automating the verification process, thereby upholding the integrity of DeFi projects.

\paragraph{For DeFi Developers}
\textit{First}, Designing a robust governance system is paramount. It is essential to align with our proposed governance taxonomy and clearly articulate the rationale behind chosen designs, especially in managing privileged functions. Implementing these designs with transparency and communicating the details to users is crucial for trust-building. 
\textit{Second}, developers must stay abreast of common governance issues and vulnerabilities. By enhancing governance mechanisms and refining implementation processes, they can address critical concerns related to ownership and tokenomics, thereby bolstering the security and reliability of their applications. Particular attention should be paid to ownership protection mechanisms, such as safeguards against unauthorized access to owner keys by team members or external threats. 
\textit{Third}, developers should commit to transparent and accurate communication with investors and users. Maintaining consistency between the project's whitepaper and actual code is essential to reduce misinformation and mitigate risks. This approach not only builds trust but also solidifies the project's credibility in the DeFi ecosystem.


\paragraph{For Investors and Users}
The insights from this study equip investors and users with the knowledge to make informed decisions in the DeFi arena. Understanding the governance issues, as categorized in our taxonomy, is key. Investors should scrutinize the governance frameworks, ownership structures, and tokenomics of DeFi projects for indicators of robustness and fairness. Crucial steps include:
\begin{itemize}
    \item Evaluating how a DeFi program manages ownership and the rationale behind these strategies.
    \item Investigating the rights granted by governance tokens within the project's structure.
    \item Assessing token distribution and privileged functions for potential unfair practices or inconsistencies with the whitepaper. 
    \item Understanding who holds the power to alter the DeFi code.
\end{itemize}
By comprehensively analyzing these elements, investors and users can discern the value and legitimacy of a DeFi application, thereby steering clear of scams. For example, the ability of a DeFi program to mint unlimited tokens or withdraw liquidity unrestrainedly may signal fraudulent intent. While there are inherent risks, a well-governed DeFi project can also present significant opportunities. Thus, a balanced approach in evaluation is crucial.


\paragraph{For Regulators and Policymakers}
Regulators and policymakers stand to gain significant insights from this research, particularly in understanding the nuanced governance challenges in DeFi. This study underscores the necessity for regulatory frameworks that go beyond assessing code vulnerabilities to encompass the entire governance structure of DeFi projects. A key area for future policy development is the role of the whitepaper; it is time to engage in discussions about its legal significance within the DeFi ecosystem. For instance, considering whether whitepapers should be subjected to regulatory oversight, given their role in outlining project fundamentals, is crucial. Additionally, governance issues often center around ownership and incentive mechanisms. Therefore, regulatory bodies should closely monitor privileged functions that have a substantial impact on DeFi tokenomics, ensuring they align with both investor protection and market integrity. While doing so, it is important to strike a balance between effective regulation and fostering an environment that encourages innovation in the DeFi space.



\section{Related Work}
\label{sec:related_work}
\paragraph{Studies on Blockchain and Decentralized Finance}
 Blockchain employs a decentralization methodology, securely storing data in a specifically structured entity known as a block.
In particular, data incorporated into the system becomes impervious to tampering. Blockchain
technology has caused profound changes and impacts on traditional Web 2.0 and is becoming a
fundamental service of the Web, leading to the emergence of Web 3.0. Blockchain has been widely used in many fields, such as cryptocurrency, financial services, games, trading systems, and IoT~\cite{zheng2018blockchain, DIFRANCESCOMAESA202099,8487348}. \citet{zheng2018blockchain} and \citet{8487348} reviewed the fundamental techniques in Blockchain such as architecture and consensus algorithms, and also discuss several blockchain applications. \citet{DIFRANCESCOMAESA202099} study non-cryptocurrency blockchain applications and practical problems they solved, indicating that blockchain is a valuable technique for the real world. Decentralized finance~(DeFi) is one main application scenario for blockchain. \citet{meyer2022decentralized} conduct the systematic literature review about DeFi at three different levels, i.e., micro, meso and macro. \citet{SHAH2023171} reviews the various types of DeFi protocol in DeFi products. \citet{10.1007/978-3-662-63958-0_20} reviews the formal methods for DeFi to ensure its correct behavior. \citet{jiang2023decentralized} investigates the DeFi running mechanisms and systemically reviews the DeFi risks. \citet{10.1145/3558535.3559780} reviews the features and security of DeFi. 

\paragraph{Blockchain and DeFi Governance}
In the realm of blockchain and DeFi governance, several studies stand out.  Allen et al.~\cite{allen2020blockchain, allen2023exchange} propose a descriptive framework to understand blockchain governance, and later develop an exchange theory for Web3 governance. Ekal et al.~\cite{ekal2022defi} bridged traditional finance governance models with blockchain-based mechanisms. Kiayias et al.\cite{kiayias2022sok} conducted a comprehensive examination of the characteristics of blockchain governance. Marc et al.~\cite{DeFieurofi} and Gogel et al.~\cite{gogel2021defi} discussed its main forms, offering insights into this emerging field.
 Liu et al.\cite{LIU2023111576} performed a systematic review analysis, summarizing the concept of blockchain governance as the system and procedures established to ensure that the development and implementation of blockchain technology complies with legal, regulatory and ethical obligations. 
 Furthermore, Liu et al.\cite{LIU2022102090} proposed a governance framework specifically for blockchain technology. Other foundational works~\cite{pelt2021defining, 10.1007/978-3-030-91983-2_6,werner2020towards} have also explored the realm of blockchain governance, offering highly abstract and conceptual insights. Another relevant study~\cite{fusco2021decentralized} looks at the DeFi revenue models and governance systems based on the analysis of real DeFi applications. Some works study governance risks and code update. Bekemeier et al.~\cite{10.1145/3510487.3510499} systematically analyzes risks in DeFi, including DeFi governance. Bhambhwani et al.~\cite{bhambhwani2022governing} analyzed the top 50 DeFi protocols and indicated potential governance risks. Erik et al.~\cite{smartgovernanceforsmartcontracts} discuss smart contract governance on smart contract upgrade in permissionless and permissioned blockchains. Reports from international organizations like the OECD and BIS~\cite{eurofi,wifpr,OECD,BIS} discuss the current state and potential impacts of DeFi, providing a broader policy perspective. All of these works do not study the governance issues in DeFi applications. We benefit from the previous studies and build a governance taxonomy to study the governance issues in DeFi.

\paragraph{DeFi Security} Since its inception, blockchain technology has held a strong affinity with finance, and as such, its security issues often precipitate substantial financial losses. For example, RONIN\footnote{https://rekt.news/ronin-rekt/} lost \$624M because the attacker found a way to access the additional validator. The increased focus on blockchain security has led to the emergence of numerous tools to identify and rectify vulnerabilities, such as Slither~\cite{feist2019slither} and ContractFuzzer~\cite{jiang2018contractfuzzer}. \citet{jcp2020019} study the 13 common vulnerabilities and compare different security tools. Nevertheless, these resources are predominantly code-centric, overlooking design-level vulnerabilities, like those that pertain to flash loan attacks. These security gaps depend mainly on human auditing for detection and correction. \citet{LI202210378} comprehensively analyze the security issues of DeFi at each blockchain layer. Since DeFi defines an economy system and interacts with the real world, its vulnerability is not only limited in its self weakness but also includes some more complex risk issues. \citet{10.1145/3368089.3409740} studies the fairness problem in DeFi. \citet{trozze2023degens} study the financial fraud in DeFi. \citet{torres2019art} study honeypot smart contracts and develop a tool to detect this scam. \citet{PonziGuard} study Ponzi scam in DeFi.

\paragraph{Code-Document Inconsistency}
Inconsistency between code and documents is the common issue in software evolution. \citet{8813274} systematically investigate the inconsistency between code and documents in a large dataset. \citet{tan2024detecting} study the outdated documents and analyze the reason why the document is not synchronized with the latest code. Recent research works use machine learning approaches to detect inconsistencies between code and documents. \citet{rabbi2020detecting} use the siamese recurrent network to detect inconsistency based on word tokens in code and documents. \citet{kim2016automatic} employ natural language processing to detect inconsistent identifiers. \citet{panthaplackel2021deep} utilize the graph neural networks to detect code-doc inconsistencies based on AST. \citet{RANI2023111515} conducted a literature review on code-document quality and found that most works focus on Java programs, resulting in poor generation performance. All of the works focus on the code function and descriptions. Our work employs the foundational model to solve the inconsistency of financial models between code and whitepapers on a fine-grained level~(expression level). 

\paragraph{NLP Topic Model and Foundation Model} The topic model~\cite{churchill2022evolution} is a useful text analysis tool and can identify the topic words in the documents without the training phase. A. Abdelrazek, Y. Eid et al.~\cite{abdelrazek2022topic} group topic molds into four categories, algebraic, fuzzy, bayesian probabilistic, and neural topic models. Since the appearance of BERT~\cite{devlin2018bert}, a large number of deep learning-based topic models have emerged~\cite{zhao2021topic}, like Sentence-BERT~\cite{reimers-2019-sentence-bert} and BERTopic~\cite{grootendorst2022bertopic}. Recently, the foundation models like ChatGPT and StarCoder~\cite{li2023starcoder} have demonstrated outstanding performance on a multitude of tasks related to documents and code. C. Zhang et al.~\cite{zhang2023one} illustrate that ChatGPT is at the initial level of general intelligence. Prompt~\cite{10.1145/3560815} technique is critical for these foundation models. P. Liu et al.~\cite{10.1145/3560815} systemically investigate prompt engineering in NLP and indicate that research on prompt theory should be enhanced.

\section{Threats to Validity}
\label{sec:threat}
In this study, there are some threats-to-validity factors that need to be considered. These threats include data bias, the subjectivity of topic models due to the limited training dataset, as well as limitations of the inconsistency detector about the code and the whitepaper.

First, this study primarily focuses on existing DeFi application projects and is limited to the analysis of governance issues. There may be selection bias in the sample. The DeFi ecosystem consists of a variety of DeFi applications, and it also is an evolving and changing field, with the possibility of new application projects and governance issues emerging. An incomplete sample set may result in biases in the analysis. It is possible that the research findings may not fully cover future trends in DeFi development without enough data. In order to address this problem, we selected up to 17 reputable Web3 security companies and collected over 4000 audit reports, aiming to include a diverse range of DeFi application projects.

Second, we filtered the data using keywords, and further analysis and summaries were conducted using the topic model, BERTopic\cite{grootendorst2022bertopic}. While this approach provides a strong analytical framework, there is still some level of subjective judgment when it comes to extracting and interpreting key themes because of the possible bias knowledge of AI models. AI models are limited by their training data. To reduce its impact, researchers conducted multiple independent analysis, and engaged in careful discussions and negotiations to reach a consensus. Additionally, the feasibility of the verification tool about the consistency between the code and the whitepaper also has some limitations. Due to variations in the formatting of different whitepapers and source code styles, the tool may encounter challenges in practical applications. The large language model training corpus includes a wide range of codebases and document styles. We chose to utilize the large language model~(ChatGPT) to reduce the impact of this diversity in styles of code and whitepapers. 



\section{Conclusion}
\label{sec:conclusion}

This paper presents a comprehensive study of governance issues in decentralized finance (DeFi) applications. Drawing on the existing research literature and industry blogs, we propose a novel DeFi governance taxonomy for governance issues, categorizing and analyzing them through the lenses of governance design and implementation. To analyze governance issues, we collected 4,446 audit reports from 17 reputable Web3 security companies, amounting to a total of 26,037 issues. To grasp the attributes of the collected data, we examined the severity levels and resolution statuses of these issues. Then, we identified in the audit reports that 7,346 issues were related to governance according to the governance taxonomy. We discovered that most of the problems are associated with ownership and incentive mechanisms. To study deeply, we employed the BERTopic tool to conduct an extensive analysis of governance issues within each category, revealing main problematic topics~(top 3). The discovery of these topics provides clearer directions and foundations to solve these governance issues. In addition, we explore the challenges posed by maintaining consistency between the code and the whitepaper in DeFi applications. Whitepapers serve as the development-team commitment to users. In our analysis of governance issues, we identified some inconsistency issues between implemented code and the whitepapers. These discrepancies can significantly impact users. To tackle this problem, we developed a governance inconsistency detector powered by AIGC~(ChatGPT) to check semantic artifacts between whitepapers and the code, and demonstrated its effectiveness by the evaluation of eight DeFi projects, providing valuable insights for the future resolution of this inconsistency issue. 

In summary, our research systematically investigates the governance issues present in DeFi applications and reveals the main concerns of DeFi governance. We propose a novel framework to understand them. \revise{Our work highlight the potential research directions about DeFi applications for software engineering researchers. As a decentralized application, DeFi emphasizes decentralization and transparency. We should study how to standardize the DeFi governance design, development and change process to make DeFi projects credible.} Through this study, we hope to help development teams of DeFi applications, DeFi users, regulatory bodies, and researchers interested in DeFi to gain a better understanding of and address governance challenges, thereby promoting the healthy development of decentralized finance.

\bibliographystyle{ACM-Reference-Format}
\bibliography{sample-base}
\end{document}